\documentclass{JAC2003}
\usepackage{graphicx}
\usepackage{booktabs}

\begin{document}
\title{LOW SENSITIVITY OPTION FOR TRANSVERSE OPTICS OF THE FLASH LINAC AT DESY}
\author{
V. Balandin, N. Golubeva\thanks{nina.golubeva@desy.de}, 
DESY, Hamburg, Germany}

\maketitle

\begin{abstract}
The aim of the FLASH linac is to create electron bunches of small
transverse emittance and high current for the free-electron laser facility 
FLASH at DESY. Available operational experience indicates that in order 
to optimize SASE signal at different wavelengths or to fine-tune the FEL wavelength, 
empirical adjustment of the machine parameters is required and, therefore, 
the sensitivity of the beamline to small changes in the beam energy and in the 
magnet settings becomes one of the important issues which affects both, the final 
performance and the reproducibility of the results after breaks in operation.
In this article the transverse optics of the FLASH beamline with low sensitivity 
to changes in beam energy and quadrupole settings is presented. This optics is 
in operation since spring 2006 and has shown a superior performance with respect 
to the previous setup of the transverse focusing.
\end{abstract}

\section{ACCELERATOR OVERVIEW}

\begin{figure*}[t]
    \centering
    \includegraphics*[width=168mm]{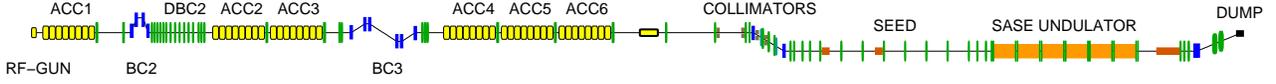}
    \caption{Current layout of the FLASH facility (top view). 
    Beam direction is from left to right. Blue, green, and orange colors mark 
    dipole and quadrupole magnets, and undulator segments, respectively.
    Yellow color indicates gun cavity, accelerating cavities and transverse-deflecting cavity.}
    \label{figOverview}
\end{figure*}

The free-electron laser FLASH at DESY, based on self-amplified spontaneous emission
(SASE), is the unique user facility operating in the VUV and the soft X-ray wavelength 
range. Since summer 2005, it provides coherent femtosecond long radiation 
to user experiments~\cite{FLASH_1, FLASH_2}. Fig.~\ref{figOverview} shows
the current layout of the facility.
Electron bunches are produced in an RF gun and accelerated in
six cryomodules (ACC1 - ACC6) each containing eight nine-cell superconducting 
cavities of the TESLA type and a doublet of superconducting quadrupoles. To achieve 
high peak current in the undulator, the bunch is longitudinally 
compressed in two magnetic chicanes. Downstream the first bunch compressor BC2 
(a four bend chicane) there is a diagnostic section (DBC2) equipped with several OTR screens 
for the measurement of beam profiles.
A second compression stage takes place after the passage through ACC2 and ACC3
and is performed using a S-type chicane (BC3). 
The last accelerating section, presently containing three cryomodules (ACC4, ACC5 and ACC6), 
accelerates the beam to the chosen final energy (presently up to 1 GeV). 
Then, the electron beam is either guided to a bypass beam line (not shown in Fig.~\ref{figOverview}) 
or brought through a collimator section to the undulator. 
The collimator section has a dogleg shape and  contains transverse and energy 
collimators for undulator protection purposes. Finally, the electron beams from the undulator 
and bypass are dumped in the same absorber (DUMP).

\section{TWISS PARAMETERS FOR ACCELERATED BEAM}

We will assume that the transverse particle motion is uncoupled
in linear approximation and will use the variables $x_{\delta}$
and $q_{\delta}$ for the description of the horizontal beam oscillations.
Here, as usual, $x_{\delta}$ is the horizontal particle coordinate and 
$q_{\delta}$ is the horizontal canonical momentum scaled with the kinetic 
momentum of the reference particle $p_{\delta}$, and 
the exact meaning of the subscript $\delta$ will be clarified in the following sections.

If the beam undergoes acceleration, $p_{\delta}$ is not constant along the accelerator
and in this case there are different methods of including the 
beam energy variation in the lattice functions which could lead, for example,
to lattice functions that, for a given periodic structure, decrease with
increasing beam energy. We will use as definition of the Twiss parameters the
second order moments of the beam distribution, namely
\begin{eqnarray}\label{TW_1}
    \beta_{\delta} = \frac{\left< x_{\delta}^2\right>}{\epsilon_x^{\delta}},
    \;\;\;\;
    \alpha_{\delta} = -\frac{\left< x_{\delta} q_{\delta}\right>}{\epsilon_x^{\delta}},
    \;\;\;\;
    \gamma_{\delta} = \frac{\left< q_{\delta}^2\right>}{\epsilon_x^{\delta}},
\end{eqnarray}
where
\begin{eqnarray}\label{TW_2}
    \epsilon_x^{\delta} = \sqrt{\left< x_{\delta}^2\right> \left< q_{\delta}^2\right>
    - \left< x_{\delta} q_{\delta}\right>^2}
\end{eqnarray}
is the rms emittance. 
With this definition, the Twiss matrix
\begin{eqnarray}\label{Equ_00}
    B_{\delta} = \left(
    \begin{array}{rr}
        \beta_{\delta}  & -\alpha_{\delta} \\
        -\alpha_{\delta} &  \gamma_{\delta}
    \end{array} 
    \right)
\end{eqnarray}
\noindent
satisfies the equation
\begin{eqnarray}\label{Equ_01}
    \frac{d B_{\delta}}{d \tau} = 
    J H_{\delta} B_{\delta} - B_{\delta} H_{\delta} J ,
\end{eqnarray}
\noindent
where $\tau$ is path length along the accelerator,
matrix $J$ is the symplectic unit and the matrix $H_{\delta}$ is always symmetric
(see equation (\ref{Equ_40}) below).

The solution of (\ref{Equ_01}) is given by the formula
\begin{eqnarray}\label{Equ_03}
    B_{\delta}(\tau) = M_{\delta}(\tau) \, B_{\delta}(0) \, M_{\delta}^{\top}(\tau) ,
\end{eqnarray}
\noindent
where the matrix $M_{\delta}$ satisfies the equation
\begin{eqnarray}\label{Equ_02}
    \frac{d M_{\delta}}{d \tau} = 
    J H_{\delta} M_{\delta}\,, \;\; M_{\delta}(0) = I \,,
\end{eqnarray}
\noindent
and is symplectic.

Alternatively, the matrix $M_{\delta}$
can be expressed using Twiss parameters, if they are known, 
in the familiar form
\begin{eqnarray}\label{Equ_04}
    M_{\delta}(\tau) = T_{\delta}^{-1}(\tau) \, R(\mu_{\delta}(\tau)) \, T_{\delta}(0) .
\end{eqnarray}
\noindent
Here $R(\mu_{\delta})$ is a 2 by 2 rotation matrix, 
$\mu_{\delta}$ is the horizontal phase advance 
(defined as usual using the integral of the reciprocal of the
betatron function) and
\begin{eqnarray}\label{Equ_05}
    T_{\delta} = 
    \left(
    \begin{array}{cc}
        1        / \sqrt{\beta_{\delta}} & 0 \\
        \alpha_{\delta} / \sqrt{\beta_{\delta}} & \sqrt{\beta_{\delta}}
    \end{array}
    \right).
\end{eqnarray}
 
Note that, if we assume that the dynamics in the variables
$x_{\delta}$ and $q_{\delta}$ is given by the matrix
$A_{\delta}$, then the following identity holds
\begin{eqnarray}\label{Equ_16} 
    M_{\delta}(\tau) = \frac{1}{\sqrt{\det\left(A_{\delta}(\tau)\right)}} \cdot A_{\delta}(\tau)\,. 
\end{eqnarray}

\section{OPTICS OPTION I}

The beam line discussed in this paper starts from the quadrupole 
doublet of the ACC1 cryomodule and ends at the entrance of the SASE undulator.
We consider neither the beam dynamics in the gun area 
nor the choice for the setting of the undulator quadrupoles (important topic,
which deserves separate consideration).

\begin{figure}[htb]
    \centering
    \includegraphics*[angle=-90,width=70mm]{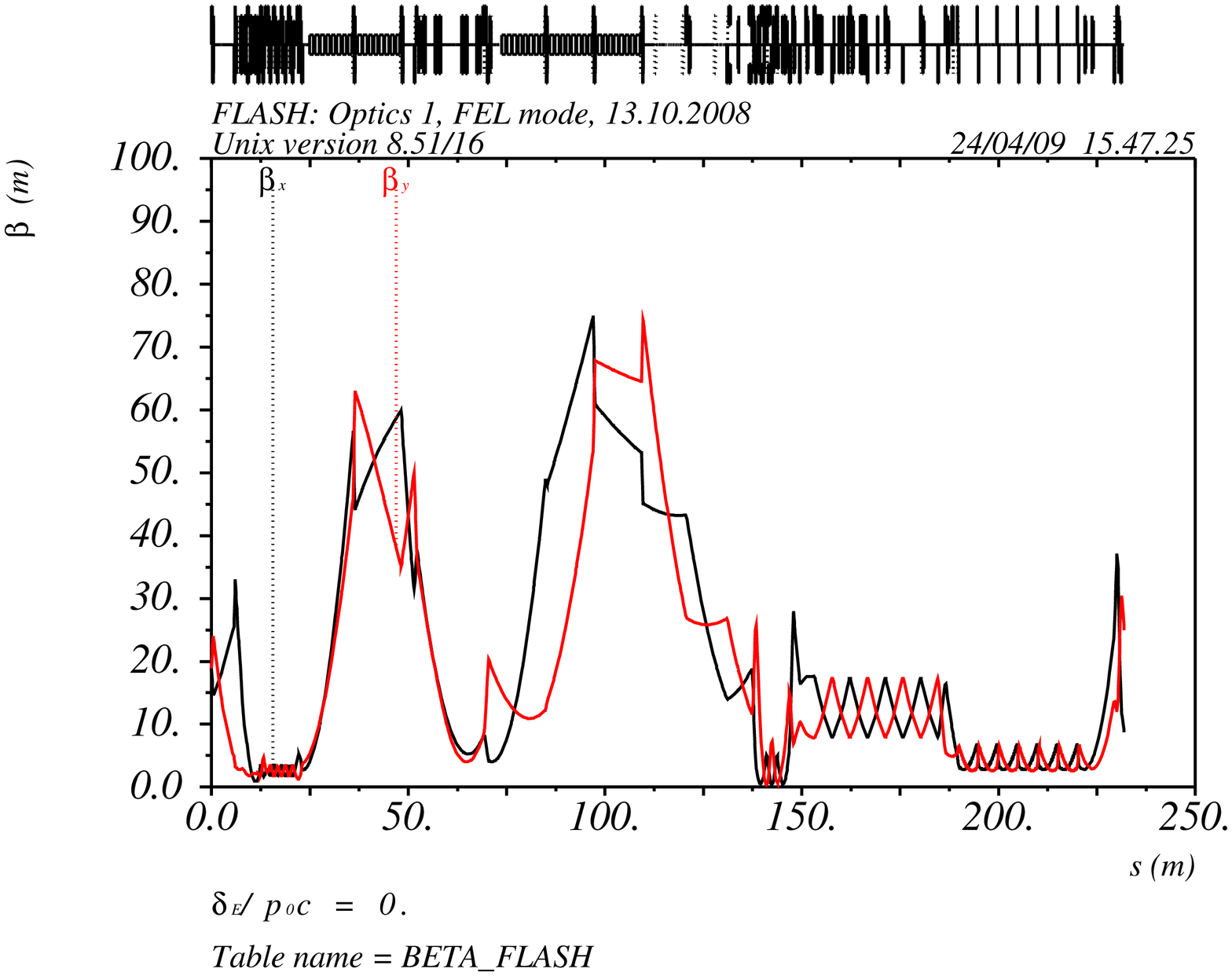}
    \caption{Optics Option I. The SASE undulator starts 
    at the position $s \approx 190 m$.}  
    \label{Op_Op_I}
\end{figure}

Optics option I, which was prepared for the 
commissioning  is described in~\cite{FLASH_4} and is shown 
in Fig.~\ref{Op_Op_I}. 
It may look somewhat non-typical for the straight beam line with its two bulges, but it was
the only way to satisfy certain lattice constraints,
which were considered at that time as important. These 
constraints include among others
\begin{Itemize}
\item  A special choice of Twiss parameters in the bunch 
       compressors BC2 and BC3 reduces the emittance growth 
       due to coherent synchrotron radiation (CSR).
\item  The selection of the optical functions in the
       collimator section is determined by the need
       to suppress the dispersion and to shape
       the beam envelop suitable for collimation purposes.
\item  Beam emittance measurements in sections DBC2 and SEED are obtained 
       from the beam sizes measured at four separated locations inside
       the FODO structure. This measurement technique provides
       its best accuracy for periodic Twiss functions and a $45^{\circ}$
       phase advance per cell.
\end{Itemize}

Optics option I was used during FLASH commissioning and during 
the first months of its operation as user facility.

\section{SENSITIVITY TO ERRORS}

Let the index $\delta = 0$ indicate the design Twiss
parameters and design setting of acceleration and focusing,
and a nonzero $\delta$ stands for the beam motion perturbed by
different imperfections.
We will characterize the cumulative effect of these imperfections
using the value of the mismatch parameter
\begin{eqnarray}\label{Equ_07}
    m_p^{\delta} = 
    \frac{1}{2} \left(
    \beta_0 \gamma_{\delta} - 2 \alpha_0 \alpha_{\delta} + \gamma_0 \beta_{\delta} \right)
\end{eqnarray}
calculated at the undulator entrance.

If we assume that $\;H_{\delta} - H_0 = O(\delta)\;$
and that the difference between design and actual Twiss parameters 
at the point $\tau = 0$ has the same order of magnitude, then
we will have to lowest order (which, in fact, is quadratic in $\delta$)
\begin{eqnarray}\label{Equ_1}
    m_p^{\delta}(\tau) = 1 + \frac{1}{2}
    \left[
    \mbox{tr}^2 \left( \Lambda(\tau) \right)
    - 4 \det \left( \Lambda(\tau) \right)
    \right] + \ldots .
\end{eqnarray}
Here
\begin{eqnarray}\label{Equ_2}
    \Lambda(\tau) = T_0^{-\top}(0) \left(
    \hat{T}(0) - \hat{H}(\tau)
    \right) T_0^{-1}(0) ,
\end{eqnarray}
\begin{eqnarray}\label{Equ_3}
    \hat{H}(\tau) = \int_0^{\tau}
    M_0^{\top}(\xi) \left( H_{\delta}(\xi) - H_0(\xi) \right) M_0(\xi) d\xi ,
\end{eqnarray}
\begin{eqnarray}\label{Equ_4}
    \hat{T} = -\frac{1}{2 \beta_0}
    \left(
    \begin{array}{cc}
        2 (\alpha_{\delta} - \alpha_0) & \beta_{\delta} - \beta_0 \\
        \beta_{\delta} - \beta_0       & 0
    \end{array}
    \right) .
\end{eqnarray}
The matrix $\hat{T}$ represents the contribution of initial beam
mismatch and the matrix $\hat{H}$ adds the mismatch accumulated due
to imperfections in acceleration and focusing.
 
In the interesting case for us, when
focusing and acceleration is provided by
quadrupoles and rotationally symmetric cavities,
the matrix $H_{\delta}$ can be expressed using the speed
of light approximation as follows
\begin{eqnarray}\label{Equ_40}
    H_{\delta} = 
    \left(
    \begin{array}{cc}
        \frac{1}{2 p_{\delta}}\frac{d^2 p_{\delta}}{d \tau^2} + k_{\delta} & 
        \frac{1}{2 p_{\delta}}\frac{d p_{\delta}}{d \tau} \\
        \frac{1}{2 p_{\delta}}\frac{d p_{\delta}}{d \tau} &  1 
    \end{array}
    \right) .
\end{eqnarray}
Here $k_{\delta}$ is the quadrupole coefficient and the focusing effect of 
accelerating field is completely
hidden in dynamics of the reference momentum $p_{\delta}$.

Let us make more simplifications and, first, assume that the effect of
an initial beam mismatch is smaller than the cumulative effect of
focusing perturbations (the procedure of matching the beam parameters at
the entrance of the bunch compressor BC2 is well established at FLASH), and, second,
neglect in the matrix $H_{\delta} - H_0$ the difference in
RF focusing due to difference in design and actual accelerating field. 
Then we have
\begin{eqnarray}\nonumber
    m_p^{\delta}(\tau) = 1 +  \frac{1}{2}
    \left[ \left(
    \int_0^{\tau} (k_{\delta} - k_0) \beta_0 \cos(2 \mu_0) d \xi
    \right)^2 +\right.
\end{eqnarray}
\begin{eqnarray}\label{Equ_654321}
    +
    \left.
    \left(
    \int_0^{\tau} (k_{\delta} - k_0) \beta_0 \sin(2 \mu_0) d \xi
    \right)^2
    \right] + \ldots .
\end{eqnarray}

As a final step, we introduce a positive weight function $g_0$
and obtain an estimate

\begin{eqnarray}\label{EEEq}
    m_p^{\delta}(\tau) \leq 1 +
    \frac{1}{2}
    \left(
    \int_0^{\tau} \left(g_0 \beta_0\right)^2  d \xi
    \right)
    \left(
    \int_0^{\tau} \Delta_{\delta}^2  d \xi
    \right) ,
\end{eqnarray}
where $\;\Delta_{\delta} = (k_{\delta} - k_0)\, /\, g_0$.

As criteria for the optics sensitivity to quadrupole errors
we choose the value
\begin{eqnarray}\label{EEEq_2}
    \sqrt{\int_0^{\tau} \left(g_0(\xi) \beta_0(\xi)\right)^2  d \xi} .
\end{eqnarray}

The purpose of the weight function is to 
make the values of $\Delta_{\delta}$ approximately equal
in order of magnitude for all quadrupoles and , therefore, it 
reflects our knowledge or our hypothesis  about the error distribution,
and is, eventually, a designer choice.

\section{OPTICS OPTION II}

\begin{figure}[htb]
    \centering
    \includegraphics*[angle=-90,width=70mm]{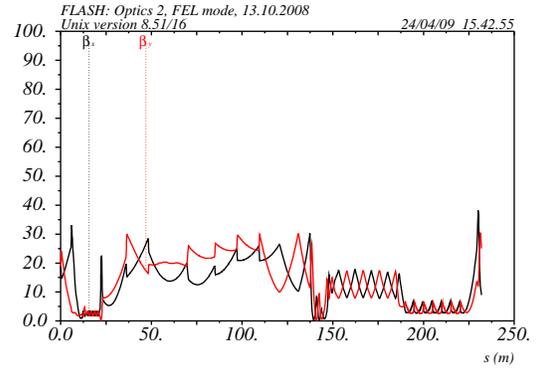}
    \caption{Optics Option II.}
    \label{Op_Op_II}
\end{figure}

In the design of the optics option II we used $k_0$ as a weight functions,
i.e. we placed the main attention to the relative errors in
the quadrupole $k$-values. This was motivated not only by possible
uncertainties in the acceleration field distribution along
the accelerating modules, but also by uncertainties in knowledge
of the transfer coefficients between quadrupole fields and power
supply currents.  

This optics, which is in operation since spring 2006 and is 
shown in Fig.~\ref{Op_Op_II}, reduces the sensitivity to quadrupole errors
by a factor of two as compared with the previous optics using criteria (\ref{EEEq_2}), 
but does not show the special behavior of the beta functions in the
bunch compressor BC3 and moderately changes the beta functions
through the collimator section.

\section{CURRENT FLASH OPTICS}

\begin{figure}[htb]
    \centering
    \includegraphics*[angle=-90,width=70mm]{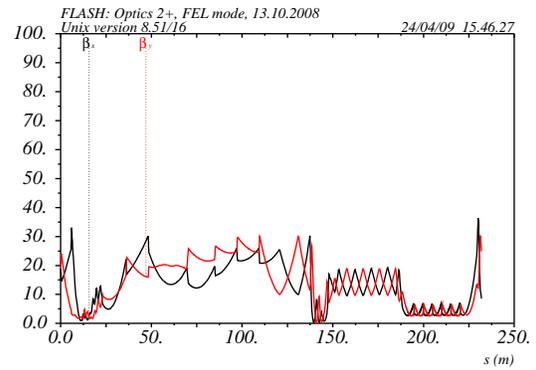}
    \caption{Current FLASH Optics (Optics Option II+).}
    \label{Op_Op_IIP}
\end{figure}

With the operational experience gained and with almost
all quadrupoles being remeasured, now we are much more
confident in the setting of transverse focusing.
Therefore optics option II was further developed and evolved
into a focusing setup which we call optics option II+, and
which is shown in Fig.~\ref{Op_Op_IIP}. This optics,
which is in operations since the beginning of 2008, 
improves the chromatic beam transfer properties of the
optics option II in the DBC2 and SEED sections.
In the same time it violates, to a greater or lesser extent, 
all constraints which were considered as important before
the beginning of operations.

\section{ACKNOWLEDGEMENT}

We thank the colleagues from the FLASH team
for their interest in our optics studies and for
their help during practical optics setup and measurements.
We are specially grateful to H. Mais for the interesting discussion 
and the careful reading of the manuscript.

\end{document}